\documentclass[useAMS,usenatbib,usegraphicx]{mn2e}

\usepackage[fleqn]{amsmath} 
\newcommand{\R}{R_\mathrm{master}}

\title[The sweeping rate in 
diffusion-mediated reactions on dust grain surfaces]{The sweeping rate in 
diffusion-mediated reactions \\ on dust grain surfaces}
\author[I.\ Lohmar and J.\ Krug]{I.\ Lohmar and J.\ Krug \\ 
Institut f\"ur theoretische Physik, Universit\"at zu K\"oln,
Z\"ulpicher Str.\ 77, 50937 K\"oln, Germany}
\begin{document}
\date{\today}
\pagerange{\pageref{firstpage}--\pageref{lastpage}} \pubyear{2002}
\maketitle

\label{firstpage}

\begin{abstract}
  A prominent chemical reaction in interstellar clouds is the
  formation of molecular hydrogen by recombination, which essentially
  takes place on dust grain surfaces.  Analytical approaches to model
  such a system have hitherto neglected the spatial aspects of the
  problem by employing a simplistic version of the sweeping rate of
  reactants. We show how these aspects can be accounted for by a
  consistent definition of the sweeping rate, and calculate it exactly
  for a spherical grain. Two regimes can be identified: Small grains,
  on which two reactants almost surely meet, and large grains, where
  this is very unlikely. We compare the true sweeping rate to the
  conventional approximation and find a characteristic reduction in
  both regimes, most pronounced for large grains. These effects can be
  understood heuristically using known results from the analysis of
  two-dimensional random walks.  We finally examine the influence of
  using the true sweeping rate in the calculation of the efficiency of
  hydrogen recombination: For fixed temperature, the efficiency can be
  reduced considerably, and relative to that, small grains gain in
  importance, but the temperature window in which recombination is
  efficient is not changed substantially.
\end{abstract}


\section{Introduction}
The complex chemistry of interstellar clouds originates from reactions
taking place both in the gas phase and on the surfaces of dust grains. The
most important surface reaction is the formation of molecular hydrogen,
which is highly inefficient in the gas phase \citep{Vidali05}. Accordingly,
there has been much recent interest in developing accurate and efficient
methods for the modelling of chemical reactions taking place on dust grain
surfaces \citep{Herbst03}. The most widespread approach for treating
networks involving both surface and gas phase reactions is based on chemical
rate equations, where the populations of different species are described by
continuous concentration variables and the rate of production of a species
is proportional to the product of the concentrations of its constituents
\citep{Pickles77}.

However, rate equations become inappropriate
for small grains subject to low fluxes of reactants, 
because the concept of a continuous concentration is
questionable when the typical number of reactants on a grain
drops below unity \citep{Charnley97,Caselli98}. In this regime
it is necessary to consider the full probability distribution
$P(N)$ of the number of reactants $N$ on the grain (say, the 
number of H atoms), either within a stochastic Monte Carlo simulation
\citep{Tielens82,Stantcheva01} or by solving the master equation governing the
time evolution of $P(N)$ \citep{Green01,Biham01,Biham02,Lipshtat04}.

None of the three modelling approaches mentioned above takes explicit account
of the process by which reactants migrate and meet on the grain surface,
through thermal hopping or quantum tunnelling. All such spatial aspects of
the problem are instead lumped into the \textit{sweeping rate} $A$, which
is usually defined as the inverse of the time required for a single
atom to visit all $S$ adsorption sites on the grain, and is taken to 
be of the form \citep{Stantcheva01}
\begin{equation}
\label{A}
A = a/S
\end{equation}
where $a$ denotes the rate of hops between neighboring adsorption
sites.  This expression is problematic for two reasons. First, since a
random walker in two dimensions occasionally revisits a site, it
requires more than $S$ hops to explore the entire surface
\citep{Montroll65,Hughes95,Krug03}. Second, and more fundamentally,
the actual process of interest -- the joint diffusion of two atoms
which terminates when either one of them desorbs, or the two meet and
react -- is distinct from, and not simply related to, the exploration
of the grain by a single atom.  To the best of our knowledge, so far
the only approach that overcomes these limitations is the fully
microscopic Monte Carlo simulation of the hopping and reaction of
molecules on the lattice of adsorption sites \citep{Chang05,Cuppen05},
which cannot be used for the modelling of large reaction networks over
long times.

The goal of this paper is to provide a consistent definition of the
true sweeping rate, and to show how it can be evaluated exactly for a
spherical grain.  In Sect.~\ref{EncounterSweeping} we establish a
transparent interpretation of the sweeping rate and express it in
terms of the encounter probability of a pair of atoms on the grain.
This encounter probability is calculated for a spherical grain in
Sect.~\ref{pencCalc}, and two regimes of parameters are identified,
for which there are simple approximate results. This section also
highlights a quantitative relation between the encounter probability
and the deviation of the distribution of the number of reactants on
the grain from a Poisson distribution. In Sect.~\ref{SweepingRate} the
results are used to obtain the true sweeping rate in terms of physical
parameters, which is then compared to the conventional approximation.
We find that the true sweeping rate is reduced considerably, and the
effect is most pronounced for large grains. Random walk considerations
are employed to reproduce asymptotically exact expressions. We then
examine the effect of using the true sweeping rate on the
recombination efficiency (as a function of grain size and of grain
temperature) in Sect.~\ref{Recombination}, for a set of physical
parameters that has been used in other work, and that enables us to
compare our results to those of microscopic Monte Carlo simulations.
A summary and our conclusions are presented in
Sect.~\ref{Conclusions}.

We focus the discussion on the specific case of hydrogen
recombination, but expect the results to be relevant also for other, more
complex diffusion-mediated reactions.

\section{Basic relations and definitions}
\label{EncounterSweeping}
We consider a grain subject to a flux $F$ of H atoms per unit time, which
can desorb at rate $W$. Once formed, molecular hydrogen is assumed to desorb
rapidly, or at least not to affect the diffusion and reaction of H atoms.
The standard rate equation governing the mean number $\langle N \rangle$ of
H atoms on the grain then reads \citep{Biham98}
\begin{equation}
\label{rate}
\frac{d \langle N \rangle}{dt} = F - W \langle N \rangle - 2 A \langle N \rangle^2.
\end{equation}
The sweeping rate $A$ appears in the last term on the right hand side,
which contains the key approximation inherent in the rate equation
method, i.e.\ setting the reaction rate proportional to the product of
the mean numbers of reactants.  Since each reaction consumes two
atoms, the recombination rate is thus taken to be
\begin{equation}
\label{rate2}
R_{\mathrm{rateeq}} = A \langle N \rangle^2. 
\end{equation}
Within the master equation treatment, this is replaced by
the expression  
\citep{Green01,Biham01,Biham02}
\begin{equation}
\label{master}
\R = A \langle N (N - 1) \rangle,
\end{equation}
which correctly accounts for the fact that the reaction
rate is proportional to the number of \textit{pairs} of atoms, and
hence vanishes when $N = 1$.
Equations (\ref{rate2}) and (\ref{master}) are seen to be identical
whenever $\langle N^2 \rangle - \langle N \rangle^2 = \langle N \rangle$, 
i.e.\ when the variance of the distribution $P(N)$ of the reactant number
equals its mean. This is a defining property of the Poisson distribution, 
which holds irrespective of the mean of the distribution. 

The comparison of (\ref{rate2}) and (\ref{master}) shows 
that a failure of the rate equation
approach, in the sense of a large discrepancy between the 
two expressions, is not necessarily implied when the 
mean number of H atoms $\langle N \rangle$ becomes small
compared to unity \citep{Biham05}. The rate equation
can be safely applied to any arbitrarily 
small region of a macroscopic surface, as long as atoms
can freely enter and exit to ensure that 
the Poisson statistics of the number of reactants in
the region is maintained. Rather, the failure of the rate
equation (\ref{rate}) on small grains is a consequence
of the \textit{confinement} of the reactants on the grain
surface, which implies that every adsorption site is visited
many times and the probability for the two reactants to 
meet is sharply enhanced. As a consequence, the probability 
to find two atoms simultaneously on the grain is strongly
reduced compared to the Poisson distribution, and 
$\R \ll R_{\mathrm{rateeq}}$. A precise
quantitative relation between the encounter probability
for two atoms and the deviation of $P(N)$ from a Poisson
distribution will be given below in Sect.~\ref{NonPoisson}.

The expression (\ref{master}) provides a transparent 
interpretation of the sweeping rate $A$: Since the number
of pairs of atoms on the grain is $N(N-1)/2$, $2A$ is the 
rate at which \textit{pairs of atoms are removed 
by the reaction}. The alternative pathway for a pair to disappear
is through the desorption of one of its constituents, which occurs
at rate $2W$. We conclude that the probability $p_{\mathrm{enc}}$ for
a pair of atoms to meet and react before one of them desorbs is
given by
\begin{equation}
\label{penc1}
p_{\mathrm{enc}} = \frac{2A}{2A + 2W} = \frac{A}{A + W}.
\end{equation}
Conversely, the relation (\ref{penc1}) can be used to 
express the sweeping rate in terms of $p_{\mathrm{enc}}$ as 
\begin{equation}
\label{Anew}
A = \frac{W p_{\mathrm{enc}}}{1 - p_{\mathrm{enc}}}.
\end{equation}
In the next section we show how $p_{\mathrm{enc}}$ can be computed
analytically in terms of the model parameters.
  
\section{Encounter probability}
\label{pencCalc}

Since the confined geometry plays an important role for the sweeping
rate, we need to specify the shape of the grain. We will treat the
simplest case of a sphere, which has the crucial advantage that all
adsorption sites are equivalent. This is in contrast e.g.\ to the disc
geometry employed previously in a similar calculation \citep{Krug03}.

The definition of the encounter probability involves a pair of atoms, each
of which hops to adjacent adsorption sites at rate $a$ and desorbs at rate
$W$.  Because of translational invariance on the surface of the sphere, this
situation is equivalent to that of a single particle moving with the double
hopping rate $2a$ and desorbing at twice the rate $2W$, while the other atom
remains fixed and present all the time. As we are only concerned with the
encounter probability (as opposed to the time of encounter), multiplying
both rates by a common factor of 2 does not change the result [compare to
(\ref{penc1})]; hence we revert to the original rates $a$ and $W$ for the
moving particle and keep the second particle fixed and present throughout.

\subsection{Stationary diffusion problem}
\label{Diffusion}

Following the approach of \citet{Krug03}, the next step consists in passing
to a continuum limit in which the occupation probability $n(\bmath{x})$ of
the moving atom satisfies a stationary diffusion equation. The spherical
grain has a radius $R$, while the fixed `target atom' is modelled as a
circular surface area of radius $r$, located at the north pole of the
sphere. Since we assume the grain to be large enough to contain at least a
few hundred adsorption sites, $r\ll R$.  In the original discrete setting
the sphere is tiled by adsorption sites forming some regular lattice. The
radius $r$ of the target atom then corresponds to a distance of $2r$ between
adjacent adsorption sites.  In the appendix~\ref{gDerivation} it is shown
for some common lattice structures that the number of adsorption sites is
given by
\begin{equation}
\label{S}
S=g(2R/r)^2,
\end{equation} 
where the factor $g=\mathcal O(1)$ reflects the lattice geometry.

The diffusion constant $D$ for a two-dimensional random walk on the lattice
of adsorption sites is related to the hopping rate $a$ by $D=a(\delta
x)^2/4$, where $\delta x = 2r$ is the length of a single step, and hence
$D=ar^2$ \citep{Michely04}.  We thus arrive at the stationary diffusion
equation
\begin{equation}
\label{Diff}
D\nabla^2 n+\frac{F}{4\pi R^2}-Wn=0
\end{equation}
describing the occupation probability $n$ of the moving atom in the steady
state of impingement, desorption and reaction. To utilise the 
azimuthal symmetry of the problem we go to spherical polar
coordinates, writing the Laplace operator in the form
\begin{equation}
\nabla^2= \frac{1}{R^2\sin\theta} \frac{\partial}{\partial \theta}
\sin\theta\ \frac{\partial}{\partial \theta}\,.
\end{equation}
The reaction removing the pair of atoms is accounted for by an absorbing
boundary condition at the boundary of the fixed atom,
\begin{equation}
\label{bc}
n(\theta=r/R)=0\,.
\end{equation}

\subsection{Exact solution}

The non-negative solution of (\ref{Diff}, \ref{bc})
that is finite at the south pole $\theta=\pi$ is
found after transforming to the new variable $z=\cos\theta$, and is given in
terms of Legendre functions of the first kind $P_\nu(z)$ 
by\footnote{Mathematical details can be found in
  \citet{Gradshteyn00}, sections 8.7 and
  8.8, especially 8.706f., 8.823 and 8.840 to 8.842. Note that, although the
  index $\nu$ can become complex, all physical expressions are real.}
\begin{equation}
n(\theta)
= \frac{F}{4\pi R^2W}
 \left[1-\frac{P_\nu(-\cos\theta)}{P_\nu(-\cos(r/R))}\right],
\end{equation}
where the index reads
\begin{equation}
\label{index}
  \nu = -\frac12+\sqrt{\frac14-\left(\frac{R}{\ell_\mathrm D}\right)^2}\,,
\end{equation}
and we have introduced the \textit{diffusion length} 
\begin{equation}
\label{lD}
\ell_\mathrm D=\sqrt{D/W}.
\end{equation}
This is the typical distance an atom on an unbounded surface would diffuse
prior to desorbing. 

The encounter probability for the pair of atoms (already averaged over
all possible starting points) then consists of two parts. The first
part is given by the fraction of all impinging atoms that reach the
target by diffusion, that is the fraction of the impingement flux $F$
that enters the target as a diffusion flux:
\begin{equation}
p_\mathrm{diff}
=2\pi R\sin(r/R) \times
\frac{D}{R} \left.\frac{\partial n}{\partial \theta}\right|_{\theta=r/R}\times
\frac{1}{F}\,.
\end{equation}
Using a well-known identity\footnote{\citet{Gradshteyn00}, equation
  8.752, 1.} this yields
\begin{equation}
\label{pdiff}
p_\mathrm{diff}
  =\frac{\sin(r/R)}{2(R/\ell_\mathrm D)^2}
  \frac{P_\nu^1(-\cos(r/R))}{P_\nu(-\cos(r/R))}\,,
\end{equation}
where $P_\nu^1 = - (1 - z^2)^{1/2} d P_\nu/dz$. However, a (small)
fraction of atoms impinges directly on top of the target area, and
these are responsible for an additional contribution to the encounter
probability, namely the (purely geometrical) ratio
$u\equiv(1-\cos(r/R))/2$ of the target versus the total surface
area.\footnote{Although this term is of the order of $(r/R)^2\ll1$, we
  found it must be accounted for in certain situations. The essential
  reason for this is that, while $u\ll p_\mathrm{diff}<p_\mathrm{enc}$
  is valid in \emph{all} regimes of interest, we will also be
  concerned with the complementary quantity $1-p_\mathrm{enc}$
  appearing in (\ref{Anew}). This however can become arbitrarily
  small, and particularly of the order of $u$ or smaller.
  Appendix~\ref{Expansion} will show that inclusion of $u$ even
  \emph{simplifies} further analysis.}

The overall encounter probability of a pair of atoms therefore is
\begin{equation}
\label{penc2}
p_\mathrm{enc}
  =\frac{\sin(r/R)}{2(R/\ell_\mathrm D)^2}
  \frac{P_\nu^1(-\cos(r/R))}{P_\nu(-\cos(r/R))}+\frac{1-\cos(r/R)}{2}\,.
\end{equation}
Equation (\ref{penc2}) is the central result of this section.  It
should be emphasised that the encounter probability is independent of
the impingement flux because it only contains information about two
atoms that are already assumed to be present on the grain. The
following section will feature the behaviour of the encounter
probability in the regimes of physical interest.

\subsection{Limiting behaviour}
\label{pencLimits}
We will first derive the approximate behaviour common to all relevant
regimes (as defined by the ordering of the three length scales, $r$,
$R$ and $\ell_\mathrm D$). It was mentioned earlier that $r\ll R$ is
necessary for the dust grain to host a reasonable number of adsorption
sites. But clearly, $r\ll\ell_\mathrm D$ as well, because otherwise
the adatom hardly performs any hops during its residence time on the
grain. In terms of the parameters of (\ref{penc2}) the common feature
of all regimes is that $r/R\ll1$, thus being the appropriate quantity
in which to expand first.

The actual calculation involves some subtleties, and we relegate most
of this to appendix~\ref{Expansion}. To leading order in $r/R$, one
obtains\footnote{\label{pencasymptotics}This approximation does not
  invalidate the reality of $p_\mathrm{enc}$ for values of $\nu$ of
  the form (\ref{index}).}
\begin{equation}
\label{p-enc-asy}
p_\mathrm{enc}^{-1}\approx\left(\frac{R}{\ell_\mathrm D}\right)^2
\left[\ln\left(\frac{2R}{r}\right)^2
-2\gamma-2\psi(\nu+1)-\pi\cot(\nu\pi)\right].
\end{equation}
Here $\psi(z) \equiv d(\ln\Gamma)/dz$, where $\Gamma(z)$ is the
$\Gamma$-function, and $\gamma=-\psi(1) \approx 0.577215665\dots$ is
Euler's constant.  To gain insight into the behaviour of the encounter
probability it is useful to examine the two remaining regimes in which
$r$ is the smallest length scale of the problem. The relation between
the grain radius $R$ and the diffusion length $\ell_\mathrm D$ conveys
a notion of `small' and `large' grains.

\subsubsection{Small grains}
Here, the ordering of lengths reads $r\ll R\ll\ell_\mathrm D$ and
hence $|\nu|\ll1$. Expanding $p_\mathrm{enc}$ about $\nu\to0-$ to
first order yields $p_\mathrm{enc}\approx 1+\nu[\ln(2R/r)^2-1]$, and
with $\nu\approx-(R/\ell_\mathrm D)^2$ we
have\footnote{\label{lnOrder}We will basically treat logarithms of
  large quantities as being of the order of unity throughout, thence
  retaining coequal numerical constants in sums, although the
  logarithms are usually significantly larger. While this might
  slightly blur the fundamental functional relation, it is numerically
  adequate in many realistic situations.}
\begin{equation}
\label{pencsmall}
  p_\mathrm{enc}\approx
 1-\left(\frac{R}{\ell_\mathrm D}\right)^2
\left[\ln\left(\frac{2R}{r}\right)^2-1\right]\to1\,.
\end{equation}
In this regime of low desorption rate each atom spends enough
time on the grain to explore all adsorption sites many times. For low
fluxes, the recombination efficiency is then limited by the rare event that
two atoms are simultaneously present on the grain. In this event however,
they almost surely meet. It is for encounter probabilities near unity that
the failure of the rate equation description is most pronounced, see
Sect.~\ref{NonPoisson}.

\subsubsection{Large grains} Now, $r\ll\ell_\mathrm D\ll R$ and
$\nu=-1/2+i\lambda$ with $\lambda\approx R/\ell_\mathrm D\gg1$. For
this form of $\nu$, the cotangent in (\ref{p-enc-asy}) is purely
imaginary and cancels the imaginary part of $\psi(\nu+1)$ (cf.\ 
footnote \ref{pencasymptotics}).  The behaviour of the
$\Gamma$-function implies $\psi(z)\approx\ln z+\mathcal O(z^{-1})$ for
arguments of large modulus, leading to
$\mathrm{Re}\,\psi(\nu+1)\approx\ln(R/\ell_\mathrm D)$ for the
remaining real part of $\psi$. Together we obtain
\begin{equation}
\label{penclarge}
p_\mathrm{enc}
\approx\left(\frac{\ell_\mathrm D}{R}\right)^2
\frac1{\ln(2\ell_\mathrm D/r)^2-2\gamma}\ll1\,.
\end{equation}
For low fluxes this means that the recombination efficiency is low
because of fast desorption, which does not allow the atoms to spend
sufficient time on the grain to react. This is the \textit{second
  order} regime of \citet{Biham02} and \citet{Krug03}, in which the
confinement of the atoms to the grain surface is not felt, and hence
the rate equation continues to apply even for small numbers of
reactants.  In both limits, (\ref{pencsmall}) and (\ref{penclarge}),
the functional dependence on the model parameters is the same as that
found earlier for the disc geometry \citep{Krug03}, taking into
account that there, logarithms of large arguments are generally
assumed large as well (dominating possible numerical constants of the
order of unity).

\subsection{Non-Poissonian statistics}
\label{NonPoisson}

In this section we derive a simple quantitative relation between
the encounter probability and the deviation of $P(N)$ from 
a Poisson distribution, in the limit where the mean number
of reactants on the grain is small ($\langle N \rangle \ll 1$).
In this regime the distribution $P(N)$ can be truncated
at $N = 2$. Hence $\langle N \rangle \approx P(1) + 
2 P(2)$ and $\langle N^2 \rangle \approx P(1) + 4 P(2)$,
which implies that the recombination rate (\ref{master})
is given by
\begin{equation}
\label{R1}
\R \approx 2 A P(2).
\end{equation}
Alternatively, $\R$ can be evaluated for 
$\langle N \rangle \ll 1$ using the 
following simple statistical argument 
\citep{Krug00,Krug03}: An atom arriving freshly on the
grain will participate in a reaction if, first, another
atom is already present [which is true with probability 
$P(1)$] and, second, if the two encounter each other
(which is true with probability $p_\mathrm{enc}$).
Multiplying this with the total flux of atoms onto
the grain, we have
\begin{equation}
\label{R2}
\R \approx F P(1) p_{\mathrm{enc}} \approx W P(1)^2  p_{\mathrm{enc}}.
\end{equation}
In the second step we have used that, to leading
order for $\langle N \rangle \to 0$, the mean
number of atoms on the grain is 
$\langle N \rangle \approx F/W \approx P(1)$.
Combining (\ref{R1}), (\ref{R2}) and (\ref{Anew})
we arrive at the central relation
\begin{equation}
\label{Poisson}
P(2) \approx \frac{1}{2} (1 - p_{\mathrm{enc}}) P(1)^2.
\end{equation}
Since for the Poisson distribution
$P(2) = P(1)^2/2$ for $\langle N \rangle \to 0$, 
we see that the factor
$1 - p_\mathrm{enc}$ is a quantitative
measure for the deviation of $P(N)$ from 
the Poisson distribution, which reflects the
depletion of the probability of pairs
caused by the recombination reaction.

The relation (\ref{Poisson}) translates
into a very simple expression for
the deviation of the rate equation
recombination rate (\ref{rate2}) from
the true rate (\ref{master}). Indeed, 
for $\langle N \rangle \to 0$ the ratio
of the two becomes
\begin{equation}
\label{rho}
\frac{\R}{R_{\mathrm{rateeq}}} 
\approx 2 P(2)/P(1)^2 \approx 1 - p_{\mathrm{enc}},
\end{equation}
which highlights once more the importance
of the confined grain geometry in the 
breakdown of the rate equation 
description. Using (\ref{pencsmall2}), the ratio
(\ref{rho}) can be directly evaluated
in terms of the physical parameters
of the system.

\section{Sweeping rate}
\label{SweepingRate}

We now translate our results back into the original language
of the discrete picture via the relations $(2R/r)^2=S/g$,
$\ell_\mathrm D/r=\sqrt{(D/r^2)/W}=\sqrt{a/W}$, and accordingly
$R/\ell_\mathrm D=\sqrt{SW/(4ga)}$. Rewriting the exact expression
(\ref{penc2}) for the encounter probability in these quantities is
trivial, but only useful for the plots to come.

\subsection{Sweeping rate for small and large grains}
\label{SweepingRateLimits}

The regime of small grains is now characterized by $SW/a\ll1$, with an
encounter probability
\begin{equation}
\label{pencsmall2}
p_\mathrm{enc}\approx1-\frac{SW}{4 g a}\left[\ln (S/g)-1\right]\,,
\end{equation}
and the sweeping rate (\ref{Anew}) becomes
\begin{equation}
\label{Asmall}
A\approx\frac{W}{1-p_\mathrm{enc}}\approx\frac{4ga/S}{\ln (S/g)-1}\,.
\end{equation}
For large grains, given by $W/a\ll1\ll SW/a$, we have
\begin{equation}
\label{penclarge2}
p_\mathrm{enc}\approx\frac{4ga}{SW}\frac1{\ln(4a/W)-2\gamma}\,,
\end{equation}
leading to a sweeping rate of
\begin{equation}
\label{Alarge}
A\approx Wp_\mathrm{enc}\approx\frac{4ga/S}{\ln(4a/W)-2\gamma}\,.
\end{equation}
Compared to the conventional approximation (\ref{A}), the sweeping
rate is seen to be reduced by a logarithmic factor in both limits (the
subtracted numerical constants in the preceding expressions will be
ignored for the remaining discussion, cf.\ footnote \ref{lnOrder}).

In both cases, the argument of the logarithm is of the order of the
number of all sites that the \textit{second} (moving) atom has visited
during its residence time.  This behaviour is to be expected from
random walk considerations for this moving atom (we still employ the
scenario of the first H atom fixed and present throughout, as
introduced at the beginning of Sect.~\ref{pencCalc}): It is a known
result for two-dimensional random walks that the mean number of
\emph{distinct} sites the atom has visited after time $t$ (and thus,
after $at$ hops on average) is (asymptotically) given by
\begin{equation}
N_\mathrm{dis}\approx \frac{\pi at}{C \ln(B at)}\,,
\end{equation}
with constants $C$ and $B$ depending on the type of lattice
\citep{Montroll65,Hughes95}. Note that this statement remains true if
$B$ is changed to \emph{any} positive value, and accordingly, we will
focus on the pre-factor depending on $C$. This constant has a value of
$C=1$ for a square, $C=\sqrt{3}/2$ for a triangular, and
$C=3\sqrt{3}/4$ for the hexagonal lattice, so that for our purposes,
we can identify (cf.\ appendix~\ref{gDerivation})
\begin{equation}
C=\pi/(4g)\,.
\end{equation}

The second ingredient of our reasoning is the effective residence time
$t_\mathrm{res}$ of the moving atom in the two limits.  If the
encounter probability is very small (large grains), the atom explores
only a small portion of the grain, and the residence time is (to first
order in $a/(SW)$) limited by desorption, $t_\mathrm{res}\approx1/W$.
The atom has therefore performed $a/W$ hops and has visited
$N_\mathrm{dis}\approx 4g (a/W)/\ln(Ba/W)$ distinct sites.  The encounter
probability (which includes an average over all spatial initial
conditions) is then given by the ratio $N_\mathrm{dis}/S$, reproducing
(\ref{penclarge2}). This is the probability for the atom to recombine
before it desorbs, which in this setting can as well be obtained as
the recombination rate $A$ times the residence time $1/W$, thereby
yielding (\ref{Alarge}) once again.

If, on the other hand, the encounter probability is near unity (small
grains), the stay of the atom is almost always ended by recombination,
and the residence time becomes $t_\mathrm{res}\approx1/A$. By the same
time, the atom has explored the entire grain, so that the number of
distinct sites visited has saturated at
$N_\mathrm{dis}(t_\mathrm{res})\approx S$.  Inverting this relation
one obtains (again to first order) $t_\mathrm{res}\approx
S\ln[SB/(4g)]/(4g a)$, thus reproducing the true sweeping rate
(\ref{Asmall}).  It is also easy to regain (\ref{pencsmall2}) without
using the general relation (\ref{penc1}): The probability for the atom
to desorb before a reaction occurs can be written as
$1-p_\mathrm{enc}$, but in this limit it is also given by the
desorption rate $W$ times the residence time $1/A$.

\subsection{Comparison with the conventional approximation}

We now want to illustrate the impact of the reduction factor that using the
true sweeping rate implies. We will use the (appropriately translated)
exact expression (\ref{penc2}) for all plots throughout.

Evaluating the reduction factor of the true sweeping rate $A$ compared to
the conventional choice $a/S$ by (\ref{Anew}), we get
\begin{equation}
\label{reduction}
\frac{A}{a/S} = \frac{SW}{a} \frac{p_{\mathrm{enc}}}{1 - p_{\mathrm{enc}}}\,.
\end{equation}
For the rest of this subsection we will not be concerned with the
influence of the lattice structure, and hence will set $g=1$. With the
translation rules put forth at the beginning of
Sect.~\ref{SweepingRate}, one can convince oneself that the only effect
that a different lattice factor $g$ has on the encounter probability
is a rescaling of the grain size $S\to S/g$. It follows that, treating
the reduction factor (\ref{reduction}) as a function of the grain
size, it is subject only to a rescaling of the variable and the value
of the function (or a shift on a logarithmic scale, respectively).

\begin{figure}
\includegraphics[width=84mm]{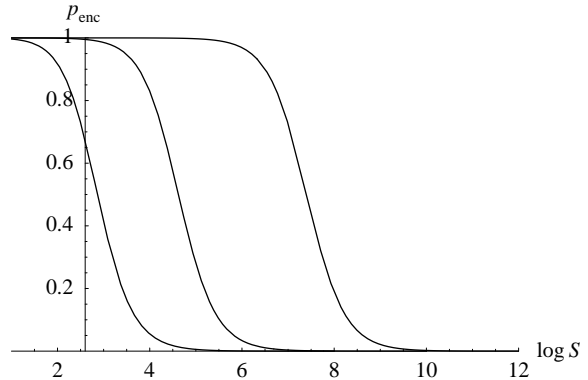}
\caption{The encounter probability $p_\mathrm{enc}$ as a function of
  the grain size $S$ (on a logarithmic scale to the basis of 10). The
  three graphs correspond (left to right) to $W/a=10^{-3}$, $10^{-5}$
  and $10^{-8}$, and the grain size discriminating between small and
  large grains is then given by the inverse $a/W$.  Starting at unity
  and monotonically decreasing, one can see the gentle roll-off of the
  encounter probability to take place around this critical size, yet
  of very similar shape in all three cases.}\label{varpencs}
\end{figure}

Fig.~\ref{varpencs} shows the encounter probability as a function of
the grain size (on a logarithmic scale to the basis of 10) for three
representative values of $W/a$. The ordinate crosses the abscissa at
$S\approx400$ (corresponding to $R/r\approx10$) so that the region to
the right of it is reasonably described by our model, cf.\ 
Sect.~\ref{Diffusion}.

\begin{figure}
\includegraphics[width=84mm]{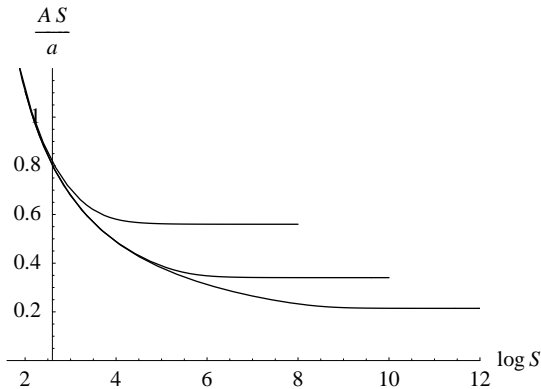}
\caption{The reduction factor of the true sweeping rate versus the
  conventional approximation, with the same conventions as in
  Fig.~\ref{varpencs} and for the same values (top to bottom) of
  $W/a$.}\label{varASdas}
\end{figure}

Fig.~\ref{varASdas} presents (with the same conventions and for the
same values of $W/a$) plots of the reduction factor $A/(a/S)$ as a
function of the grain size $S$.  The graphs converge for (absolutely)
small grains, because upon entering the small grain regime, the
different parameter $W/a$ does no longer affect $A/(a/S)$, cf.\ 
(\ref{Asmall}). With increasing grain size however, this reduction
decreases to an asymptotic value depending on $W/a$.  The
approximations given in Sect.~\ref{SweepingRateLimits} for $A$ as well
as for $p_\mathrm{enc}$ are corroborated by these two figures.

It might come as a bit of a surprise that the reduction of the
sweeping rate, while of serious account throughout, is most pronounced
for \emph{large}, not for small grains.  This can, however, easily be
understood: The key effect for the reduction of $A$ as compared to
$a/S$ is back-diffusion, the revisiting of sites by a randomly walking
atom. Small grains are those which are swept nearly entirely in most
cases. This means that back-diffusion is rather ineffective in
reducing the sweeping rate and corresponds to the observation that
different values of $W/a$ do not change the reduction factor in this
regime -- after all, recombination (not desorption) limits the
residence time, anyway. The larger the grain surface becomes, the more
does desorption compete with recombination as the limiting process.
Therefore, there are more and more initial conditions for which
back-diffusion is effective in reducing the sweeping rate. In the
limit of large grains, desorption always ends the residence time. As a
consequence, increasing $S$ no longer increases the total number of
sites an atom visits, and hence the effect of back-diffusion saturates
as well, rendering $A/(a/S)$ constant (as a function of $S$).

In Sect.~\ref{NonPoisson} it was shown that the reduction factor of
the master equation versus the rate equation recombination rate
(whenever $\langle N\rangle\ll1$) exhibits the opposite behaviour (it
is \emph{only} relevant for small grains), because this reduction is
based on a different aspect: The crucial point there was the
`non-Poissonianity' of $P(N)$ due to a large encounter probability,
and ultimately owed to the confinement of the reactants to a small
surface.

\section{Recombination efficiency}
\label{Recombination}

A quantity that is more directly connected to observations of diffuse
interstellar clouds is the recombination efficiency in the production of
molecular hydrogen.  It should therefore prove interesting to examine the
effect that the use of the reduced, true sweeping rate has on these
quantities and to compare the results to recent simulations of the process.
As the underlying analytic model we choose the master equation framework that
is now generally agreed upon. It is given by
\begin{equation}
\begin{split}
\frac{d P(N)}{dt} &= F\left[P(N-1)-P(N)\right] \\
&\quad +W\left[(N+1)P(N+1)-NP(N)\right] \\
&\quad +A\left[(N+2)(N+1)P(N+2)-N(N-1)P(N)\right]
\end{split}
\end{equation}
(with minor modifications for $N=0,\,1$).  \citet{Green01} and
\citet{Biham02} have independently found the analytic stationary solution,
yielding
\begin{equation}
\langle N\rangle = \sqrt{\frac{F}{2A}}\frac{I_{W/A}(2\sqrt{2F/A})}{I_{W/A-1}(2\sqrt{2F/A})}
\end{equation}
for the mean number of particles on the grain, and
\begin{equation}
\eta = \frac{2A}{F}\langle N(N-1)\rangle
  = \frac{I_{W/A+1}(2\sqrt{2F/A})}{I_{W/A-1}(2\sqrt{2F/A})}
\end{equation}
for the recombination efficiency $\eta=2\R/F$ with $\R$ as in
(\ref{master}). $I_\nu(z)$ are modified Bessel functions. The crucial
improvement we introduce is the consistent substitution of $A$ via
(\ref{Anew}) instead of (\ref{A}).  A convenient and physically
sensible parameter for the impingement flux consists in $f/W$, where
$f$ is the (effective) flux per site as defined by $F=fS$. Our full
set of parameters now includes $W/a$, $f/W$ and $S$.

For easiest comparison with previous work in the field, we employ the
following standard scenario: We assume desorption and hopping to be
thermally activated by the grain temperature $T$,
\begin{equation}\label{ExampleRates}
W=\nu\exp\left(-\frac{E_1}{k_\mathrm B T}\right), \qquad
a=\nu\exp\left(-\frac{E_0}{k_\mathrm B T}\right),
\end{equation}
with a uniform attempt frequency of $\nu=10^{12}\,\mathrm s^{-1}$. In an
analysis of temperature-programmed desorption experiments \citet{Katz99}
found the activation energies for an olivine surface to be $E_0/k_\mathrm
B=287\,$K and $E_1/k_\mathrm B=373\,$K.  The situation in an interstellar
gas cloud with H atoms in the gas phase at a temperature of $100\,$K is
adopted from \citet{Biham01} and leads to an effective flux of
$f=1.8\times10^{-9}\,\mathrm s^{-1}$. Altogether the physical parameters are
thereby given as
\begin{equation}
\label{PlotParams}
\frac{W}{a}=\exp\left(-\frac{86\,\mathrm K}{T}\right), \quad
\frac{f}{W}=\exp\left(\frac{373\,\mathrm K}{T}-47.665\right).
\end{equation}

\subsection{Size dependence}
\begin{figure}
\includegraphics[width=84mm]{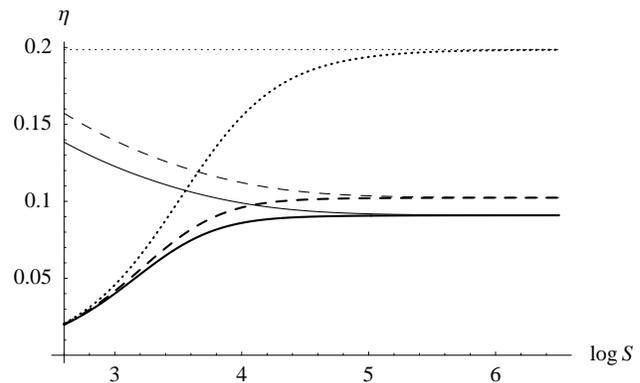}
\caption{Recombination efficiency $\eta$ as a function of the grain
  size $S$ on a $10\,$K olivine grain in the standard scenario. The
  thick lines show $\eta(S)$ as computed with the master equation and
  the true sweeping rate, the continuous line standing for a square,
  the dashed one for a triangular lattice. The thick dotted graph was
  plotted using $a/S$ as in \citet{Biham01}.  The corresponding thin
  lines show the predictions of the rate equation treatment for the
  respective choice of sweeping rate.}\label{etaS10K}
\end{figure}

We will first treat the size dependence of the recombination
efficiency $\eta$ for given temperature $T=10\,$K of the grain.  One
can see in Fig.~\ref{etaS10K} that, first, the true sweeping rate
leads to a reduced recombination efficiency as compared to the
conventional approximation in any case. Second, the quick decrease in
the efficiency as one passes from large grains (in terms of $a/W$) to
smaller sizes is shifted to the left, meaning that now, relative to
the overall efficiency, smaller grains have gained in importance. For
the sake of completeness we have further plotted the efficiency as
predicted by the rate equation treatment, i.e.\
$\eta_\mathrm{rateeq}=2R_\mathrm{rateeq}/F$ with $R_\mathrm{rateeq}$
given by (\ref{rate2}), and $\langle N\rangle$ obtained as the
stationary solution of (\ref{rate}). Clearly, using the conventional
approximation of the sweeping rate then eliminates all dependence of
$\eta_\mathrm{rateeq}$ on the size of the grain. Choosing the true
sweeping rate however results in the smaller grains even surpassing
the larger ones in recombination efficiency, as the size dependence is
then solely based on that of the reduction factor $A/(a/S)$, cf.\
Fig.~\ref{varASdas}.  This illustrates once more that the replacement
of the approximate by the true sweeping rate accounts for a different
aspect of the problem (and is of most concern in a different regime)
than the transition from the rate equation to the master equation
treatment, as we already mentioned at the end of
Sect.~\ref{SweepingRate}.

Finally, one can compare our plots with data obtained in simulations
performed by \citet{Chang05}, using a continuous-time random walk
algorithm. The efficiency we obtain for a square lattice is reasonably
close to the results of their simulations, which backs their assertion
that indeed it is back-diffusion that significantly undermines the
recombination efficiency.  However, our approach does not reproduce
the striking difference between a square and a triangular lattice in
the way that the simulations have shown (with the triangular lattice
giving rise to much more effective recombination).

The origin of this discrepancy is not clear to us at present.
However, we would like to rule out the possibility that it is an
inherent shortcoming of the continuum approximation. While it is true
that the transition from a lattice random walk to a continuous
diffusion picture obliterates the fundamental differences of lattice
types, many properties of a random walk are immune to this process.
In particular, we have seen in Sect.~\ref{SweepingRateLimits} that the
effect of the lattice type that is most important to us (the different
number of distinct visited sites) is accurately represented by the
lattice factor $g$ in our model.

As we argued there, the large grain value for the true sweeping rate
should be given by $A\approx W N_\mathrm{dis}/S$, and the residence
time is $t_\mathrm{res}\approx1/W$. It follows immediately that the
asymptotic value of $A$ for a triangular lattice should be larger than
that for a square lattice by a factor of $2/\sqrt{3}\approx1.15$,
which is the ratio of the respective values of $g$ or $C^{-1}$.  Since
for large enough grains the solution of the master equation yields
$\eta\approx2AF/W^2$ \citep{Biham02}, this factor should reappear in
the efficiencies, which is in good agreement with Fig.~\ref{etaS10K}.
Judging from the asymptotic values for large grains, the marginal
difference in recombination efficiency between the square and the
triangular lattice therefore is a reasonable outcome. Genuine lattice
random walks cannot be expected to yield a significantly larger
relative difference than our model predicts.

\subsection{Temperature dependence}
The temperature dependence of the recombination efficiency is another
interesting aspect: Ultimately, from an astrophysical point of view
the puzzle of hydrogen recombination in interstellar dust clouds is
about the temperature window in which this process is efficient.
\citet{Chang05} found that their simulations gave very similar results
compared to \citet{Biham01} in the efficiency peak, while for higher
temperatures the efficiency was somewhat smaller, and again differs
for the different lattice types (both effects as expected in our model
as well, judging from the fixed-temperature plot in
Fig.~\ref{etaS10K}).  We focus here on the comparison of the analytical
results for the different sweeping rate expressions.

It should be noted first of all that we do \emph{not} include
the mechanism of Langmuir-Hinshelwood (LH) rejection.  This mechanism
repels H atoms that impinge on occupied adsorption sites on the grain
surface and is responsible for a quick decay of the recombination
efficiency for very low temperatures, where desorption is heavily
suppressed and the H (and H$_2$) coverage on the grain is large.

As we neglect this effect and treat a fixed grain size (and therefore a
fixed impingement flux), the remaining factors governing the recombination
efficiency are the relation between the fixed $S$ and the
temperature-dependent $W/a$, and the average number of reactants on the
grain, which is smaller than, but of the order of $F/W$. As the temperature
increases, $W/a$ increases as well and we approach the regime of large grains
(cf.\ Sect.~\ref{SweepingRateLimits}) with the resulting small encounter
probability. Combined with the decreasing $F/W$ this prohibits efficient
recombination. For lower temperatures we head towards the regime of small
grains, where an encounter probability near unity and an increasing number
of reactants predict a large recombination efficiency. No finite temperature
completely prohibits diffusion; and due to the larger activation energy,
desorption is suppressed even more. Diffusion will almost always let two
atoms meet once they are on the surface, and the extremely long time this
might take at very low temperatures is a hidden feature when treating
steady-state conditions.

LH rejection should be negligible for temperatures roughly at or
larger than the peak efficiency temperature \citep{Biham01}, so we can
expect our results to validly show the breakdown for higher
temperatures. A rough estimate for a temperature below which our
results fail can be obtained using (\ref{PlotParams}). The mean
coverage of H atoms on the grain is of the order of $f/W$. The
temperature at which this coverage becomes of the order of unity
surely is beyond the validity of our assumptions, and this temperature
is given by $T_\mathrm{LH}\approx7.8\,$K.

\begin{figure}
\includegraphics[width=84mm]{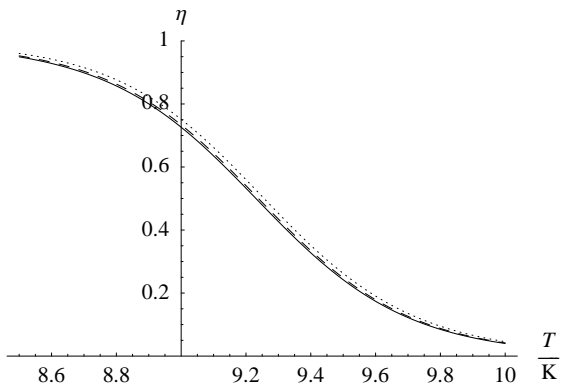}
\caption{Recombination efficiency as a function of the temperature
  $T$, for a grain of $S=10^3$ sites. Again, the continuous line shows
  the master equation result with the true sweeping rate and a square
  lattice, the dashed line represents the result for a triangular
  lattice, and the dotted line uses the conventional
  approximation. Note that this plot is horizontally stretched to
  uncover the small differences. The regime of large grains only
  starts at higher temperatures outside the plotted
  region.}\label{etaTsmallS}
\end{figure}

\begin{figure}
\includegraphics[width=84mm]{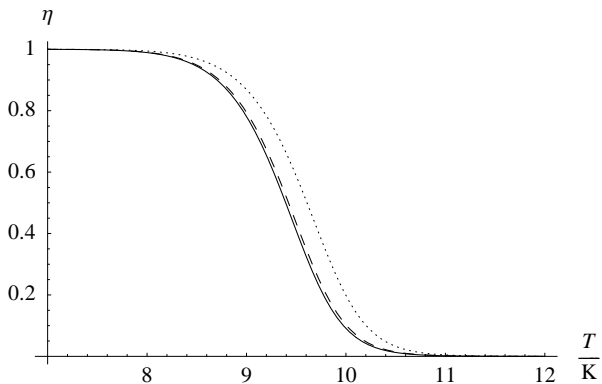}
\caption{The same plot as in Fig.~\ref{etaTsmallS}, now for a grain
  of size $S=10^6$ sites. This size is `large' for practically all shown
  temperatures.}\label{etaTlargeS}
\end{figure}

For a rather small grain of 1000 adsorption sites
(Fig.~\ref{etaTsmallS}), differences in the various predictions of the
recombination efficiency are very small, and we increased the
resolution by restriction to a very narrow temperature range. The
above estimate suggests that the breakdown of $\eta$ for $T>8\,$K
should also apply to an analogous model incorporating LH rejection. On
a larger grain with $10^6$ adsorption sites (Fig.~\ref{etaTlargeS}) one
can see that recombination ceases to be effective at a lower
temperature when employing the true sweeping rate rather than the
conventional approximation (differences between the lattice types are
still hardly perceptible).  That this effect only occurs for larger
grains is explicable with the help of Fig.~\ref{etaS10K}, which showed
that outside the efficiency peak, the reduction (of $A$ with respect
to $a/S$) is most pronounced for large grains, since only then
back-diffusion becomes effective. On the other hand our earlier
explanations imply that (independent of the fixed grain size) as we
increase or decrease the temperature, the efficiency has to approach
the limits zero and unity, respectively. Using the conventional
approximation $A=a/S$ in (\ref{penc1}) yields the quantity
$1/(1+SW/a)$ as a substitute for the encounter probability, which
shares the limiting behaviour of $p_\mathrm{enc}$ used in the argument.
Hence in both limits, the efficiency is the same for the two sweeping
rate concepts.  A fairly general suggestion of the figures is that
grain size hardly seems to matter for the upper bound of the
temperature range of effective recombination: A difference of three
orders of magnitude of the number of adsorption sites results in only
a minute change of the upper temperature bound.

\begin{figure}
\includegraphics[width=84mm]{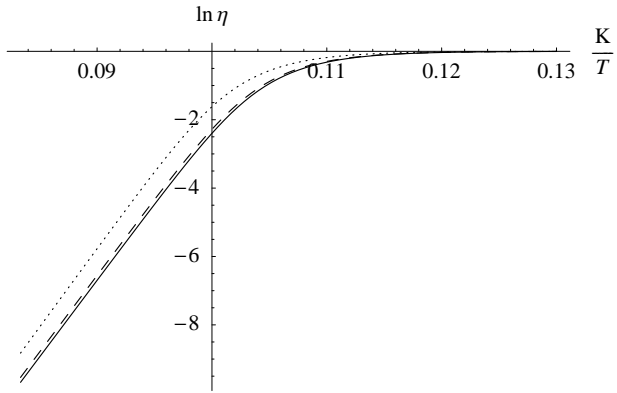}
\caption{An Arrhenius plot of the recombination efficiency under the
previous conditions.}\label{lnetaTm1}
\end{figure}

Finally, we re-plot the last graph in the Arrhenius fashion, i.e.\ as
$\ln\eta(1/T)$, which is shown in Fig.~\ref{lnetaTm1}. For rising
temperature (in the left part of the plot) there exists an effective
(negative) activation energy, which can be read off to be
$E_{\mathrm{act}}/k_\mathrm{B} \approx-440\,$K. Using the large-grain
value $\eta\approx2AF/W^2$ together with (\ref{Alarge}) and
(\ref{ExampleRates}) yields the prediction $(E_0-2E_1)/k_\mathrm
B=-459\,$K for this energy, in good accordance with the estimate.

For this astrophysical setting, we may sum up the effect of employing
the true sweeping rate (\ref{Anew}) instead of (\ref{A}) as follows:
At a given temperature, relative to the decreased overall efficiency,
smaller grains gain in importance for the recombination process, since
larger grains are more affected by back-diffusion. With a fixed grain
size, the correction is only pronounced for (absolutely) large grains,
but does not change the upper temperature bound of efficient
recombination considerably.  This bound rather seems very sensitive to
a more detailed modelling of the complex surface structure
\citep{Chang05,Perets06}.  Our results thus strengthen the claim that
hydrogen formation on dust grains crucially depends on the grain
surface structure.

\section{Conclusions}
\label{Conclusions}
We could show that it is possible to rigorously define the true
sweeping rate for diffusion-mediated reactions on a dust grain
surface, and for a spherical homogeneous grain, it could be evaluated
exactly.  The primary effect of the spatial structure of the problem
consists in a reduction of the true sweeping rate compared to the
conventional approximation, which is due to the back-diffusion of
randomly walking atoms. This reduction is moderate and dependent on
the size of the grain for small grains, whereas for increasing grain
size it grows stronger before saturating at a value dependent on the
ratio of the rates of hopping and desorption of reactants.

The picture of a spherical, homogeneous grain is highly idealized, as
is the assumption of a single reaction type, but since the effects we
have found have explanations that remain reasonable for any type of
grain and with a very general reaction network, we expect the results
to remain relevant as well.

Applying the model specifically to the recombination of hydrogen atoms
on interstellar dust grains, we could compare the recombination
efficiency that our model predicts with the results of microscopic
Monte Carlo simulations. We found that they coincide reasonably well
for a square lattice of adsorption sites, but that we cannot
reproduce, and have no explanation for, the substantially increased
efficiency on an (otherwise equal) triangular lattice simulations have
shown.

In the wake of the decreased true sweeping rate (compared to the
conventional approximation), the efficiency is reduced as well. Least
affected by this reduction, small grains become more important for the
recombination process. For a given size of the grain however, the
upper temperature bound of efficient formation of molecular hydrogen
is hardly changed.

For an improved quantitative analysis, the complex surface structure
of interstellar dust grains seems to be of paramount importance.  It
might be a worthwhile enterprise to consider more complicated
analytical models which can incorporate some of these aspects. A
possible next step would be the replacement of a single binding energy
and diffusion barrier by energy distributions, accounting for the
inhomogeneity of the lattice of adsorption sites on the grain, as has
been done in simulations \citep{Chang05}. Another interesting approach
focuses on the effect of the porosity of the dust grains
\citep{Perets06}.

\section*{Acknowledgments}
We thank E.\ Herbst for useful discussions and Q.\ Chang for
correspondence.

\appendix

\section{Lattice geometry}
\label{gDerivation}
Assuming $S$ to be large, so that curvature effects do not play any
role for the single atomic `tile', we can cover a surface area $4\pi
R^2$ by simply dividing it into such tiles, each of the area that a
single atom occupies. This tile area depends on the distance of (the
centers of) nearest neighbor sites, $2r$, and the lattice type.
Elementary trigonometry shows that, for a hexagonal lattice (which by
common terminology refers to a `honeycomb' lattice formed by the
adsorption sites, with coordination number 3), it is given by
$3\sqrt{3}r^2$, for a square lattice it is obviously $4r^2$, and a
triangular lattice (i.e.\ one with coordination number 6) has a tile
area of $2\sqrt{3}r^2$. So we have $S=g(2R/r)^2$ with a factor
$g=\pi/(3\sqrt3)$ for the hexagonal, $\pi/4$ for the square, and
$\pi/(2\sqrt3)$ for the triangular lattice, which is smaller than but
of the order of unity in all three cases.

\section{Asymptotic expansion}
\label{Expansion}
Asymptotic expansions of the Legendre functions near their singularity
are given as equations 3.9.1 (13)ff.\ in \citet{Erdelyi53}. Since it
was not clear from the outset that the degree of these expansions
sufficed, we extended the expansion with the aid of computer algebra,
and en route found a misprint (not accounted for in the errata) in
equation (15) \emph{ibid.}, which erroneously lacks a factor of 2
preceding Euler's constant, cf.\ (\ref{num})f.\ below.

We rewrite (\ref{penc2}) as
\begin{equation}
p_\mathrm{enc} = \left\{ \frac{1-z^2}{2\nu(\nu+1)}
\frac{d}{dz}\left[\ln P_\nu(z)\right]+\frac{1+z}{2}\right\}_{z=-\cos(r/R)},
\end{equation}
and then use \textsf{Mathematica} to expand the encounter probability
to first order in $(1+z)$ about $0$, keeping $\nu$ fixed.  To cast the
result into a reasonable form, we manually apply various
transformations (valid for arbitrary complex $\nu$) to get rid of all
$\Gamma$ and most $\psi$ functions, viz.\
$\Gamma(1\mp\nu) = \pm\pi/\left[\sin(\nu\pi)\Gamma(\pm\nu)\right]$,
$\psi(\pm\nu) = \psi(1\pm\nu)\mp1/\nu$, and
$\psi(1\mp\nu) = \psi(\pm\nu)\pm\pi\cot(\nu\pi)$.

Now we identify $u=(1+z)/2$, and with the definition
\begin{equation}
\label{num}
X=\ln u^{-1}-2\gamma-2\psi(\nu+1)-\pi\cot(\nu\pi)
\end{equation}
[the square brackets in (\ref{p-enc-asy})] we obtain
\begin{equation}
p_\mathrm{enc} \approx
\frac{1}{-\nu(\nu+1)X}
+\frac{1+2\nu(\nu+1)+2\nu(\nu+1)X}{-\nu(\nu+1)X^2}\cdot u
\end{equation}
as the expansion of the encounter probability up to linear order in
$u\approx[r/(2R)]^2$. The zeroth order term is equivalent to equation
(\ref{p-enc-asy}) and describes a non-trivial behaviour that we examine
in the main text. The remaining question is why the next order is not
needed in either of the two regimes discussed in
Sect.~\ref{pencLimits}. As in the main text, we regard the $\ln
u^{-1}$ term as being of the order of unity (cf.\ footnote
\ref{lnOrder}).

For large grains, there is no need to include terms of $\mathcal
O(u)$: $X$ then is of $\mathcal O(1)$ as shown in the main text, while
$-\nu(\nu+1)=(R/\ell_\mathrm D)^2\gg1$. The largest contribution of
the second term thence has an additional factor of
$-\nu(\nu+1)u\approx[r/(2\ell_\mathrm D)]^2\ll1$ compared to the first
summand.

The small grain regime requires a bit more effort, since $\nu$ and $u$
can generally be of comparable order now. Expanding about $\nu\to0-$,
we have $\psi(1+\nu)=-\gamma+\mathcal O(\nu)$ so that $X=\ln
u^{-1}-1/\nu+\mathcal O(\nu)$. Hence $-\nu X=1-\nu\ln u^{-1}+\mathcal
O(\nu^2)$ and $(\nu+1)X=-1/\nu+\ln u^{-1}-1+\mathcal
O(\nu)=-1/\nu[1-\nu(\ln u^{-1}-1)+\mathcal O(\nu^2)]$. The $\mathcal
O(u^0)$ term thus reads $1+\nu(\ln u^{-1}-1)+\mathcal O(\nu^2)$.
Similarly evaluating the $\mathcal O(u^1)$ terms yields the pre-factor
$\nu[1-\nu+\mathcal O(\nu^2)]$: Without a contribution of $\mathcal
O(\nu^0u^1)$ present, this is but a higher order correction to the
first summand.  One can further see that the direct impingement term
$u$ included in (\ref{penc2}) is of precisely that order, and thence
must have gone to cancel a corresponding term of the expression
(\ref{pdiff}).  Altogether we have shown that (\ref{p-enc-asy}) of the
main text contains all necessary terms to consistently perform the
subsequent approximations.

\label{lastpage}


\begin{thebibliography}{99}
\bibitem[\protect\citeauthoryear{Biham et al.}{1998}]{Biham98} Biham
O., Furman I., Katz, N., Pirronello V., Vidali G., 1998, MNRAS, 296, 869
\bibitem[\protect\citeauthoryear{Biham et al.}{2001}]{Biham01} Biham
O., Furman I., Pirronello V., Vidali G., 2001, ApJ, 553, 595
\bibitem[\protect\citeauthoryear{Biham \& Lipshtat}{2002}]{Biham02} Biham
O., Lipshtat A., 2002, Phys. Rev. E, 66, 056103
\bibitem[\protect\citeauthoryear{Biham et al.}{2005}]{Biham05} Biham
O., Krug J., Lipshtat A., Michely T., 2005, Small, 1, 502
\bibitem[\protect\citeauthoryear{Caselli et al.}{1998}]{Caselli98} Caselli
P., Hasegawa T.I., Herbst E., 1998, ApJ, 495, 309
\bibitem[\protect\citeauthoryear{Chang et al.}{2005}]{Chang05} Chang
Q., Cuppen H.M., Herbst E., 2005, A\&A, 434, 599
\bibitem[\protect\citeauthoryear{Charnley et al.}{1997}]{Charnley97} 
Charnley S.B., Tielens A.G.G.M., Rodgers S.D., 1997, ApJ, 482, L203
\bibitem[\protect\citeauthoryear{Cuppen \& Herbst}{2005}]{Cuppen05} 
Cuppen H.M., Herbst E., 2005, MNRAS, 361, 565
\bibitem[\protect\citeauthoryear{Erd\'elyi}{1953}]{Erdelyi53} 
Erd\'elyi A., ed., 1953, Higher Transcendental Functions, vol.\ 1.
McGraw-Hill, New York
\bibitem[\protect\citeauthoryear{Gradshteyn \& Ryzhik}{2000}]{Gradshteyn00} 
Gradshteyn I.S., Ryzhik I.M., 2000, Table of Integrals, Series, and
Products, 6th edn. Academic Press, San Diego
\bibitem[\protect\citeauthoryear{Green et al.}{2001}]{Green01} Green
N.J.B., Toniazzo T., Pilling M.J., Ruffle D.P., Bell N., Hartquist T.W.,
2001, A\&A, 375, 1111
\bibitem[\protect\citeauthoryear{Herbst \& Shematovich}{2003}]{Herbst03} Herbst E., 
Shematovich V.I., 2003, Ap\&SS, 285, 725
\bibitem[\protect\citeauthoryear{Hughes}{1995}]{Hughes95} Hughes B.D.,
  1995, Random Walks and Random Environments, vol.\ 1. Oxford
  University Press, Oxford
\bibitem[\protect\citeauthoryear{Katz et al.}{1999}]{Katz99} Katz N., Furman
  I., Biham O., Pirronello V., Vidali G., 1999, ApJ, 522, 305
\bibitem[\protect\citeauthoryear{Krug et al.}{2000}]{Krug00} Krug J., 
Politi P., Michely T., 2000, Phys. Rev. B, 61, 14037
\bibitem[\protect\citeauthoryear{Krug}{2003}]{Krug03} Krug J., 2003, Phys. Rev. E, 67, 065102(R)
\bibitem[\protect\citeauthoryear{Lipshtat \& Biham}{2004}]{Lipshtat04} Lipshtat A., 
Biham O., 2004, Phys. Rev. Lett., 93, 170601
\bibitem[\protect\citeauthoryear{Michely \& Krug}{2004}]{Michely04} 
Michely T., Krug J., 2004, Islands, Mounds and Atoms. Patterns and Processes
in Crystal Growth Far from Equilibrium. Springer, Berlin
\bibitem[\protect\citeauthoryear{Montroll \& Weiss}{1965}]{Montroll65}
Montroll E.W., Weiss G.H., 1965, J. Math. Phys., 6, 167
\bibitem[\protect\citeauthoryear{Perets \& Biham}{2006}]{Perets06}
Perets H.B., Biham O., 2006, MNRAS, 365, 801
\bibitem[\protect\citeauthoryear{Pickles \& Williams}{1977}]{Pickles77} Pickles J.B., 
Williams D.A., 1977, Ap\&SS, 52, 443
\bibitem[\protect\citeauthoryear{Stantcheva et al.}{2001}]{Stantcheva01} Stantcheva T.,
Caselli P., Herbst E., 2001, A \& A, 375, 673
\bibitem[\protect\citeauthoryear{Tielens \& Hagen}{1982}]{Tielens82} Tielens A.G.G.M., 
Hagen W., 1982, A\&A, 114, 245
\bibitem[\protect\citeauthoryear{Vidali et al.}{2005}]{Vidali05} Vidali G., Roser J., 
Manic\'o G., Pirronello V., Perets H.B., Biham O., 2005,  
J. Phys.: Conf. Ser., 6, 36


\end{thebibliography}
\end{document}